\documentclass[runningheads]{llncs}
\usepackage{graphicx}
\usepackage[utf8]{inputenc} 
\usepackage[T1]{fontenc}    
\usepackage{hyperref}       
\usepackage{url}            
\usepackage{booktabs}       
\usepackage{amsfonts}       
\usepackage{nicefrac}       
\usepackage{microtype}      
\usepackage{graphicx}
\usepackage{subfigure}
\usepackage{amsmath,amsfonts,amssymb}
\usepackage{mathtools}
\usepackage{bbm}
\newcommand{\norm}[2]{\left\lVert#1\right\rVert_{#2}}
\begin{document}
\title{Accurate Protein Structure Prediction by Embeddings and Deep Learning Representations}
%
%
\author{Iddo Drori\inst{1,2} \and
Darshan Thaker\inst{1} \and
Arjun Srivatsa\inst{1} \and
Daniel Jeong\inst{1} \and
Yueqi Wang\inst{1} \and
Linyong Nan\inst{1} \and
Fan Wu\inst{1} \and
Dimitri Leggas\inst{1} \and
Jinhao Lei\inst{1} \and
Weiyi Lu\inst{1} \and
Weilong Fu\inst{1} \and
Yuan Gao\inst{1} \and
Sashank Karri\inst{1} \and
Anand Kannan\inst{1} \and
Antonio Moretti\inst{1} \and\\
Mohammed AlQuraishi\inst{3} \and
Chen Keasar\inst{4} \and
Itsik Pe'er\inst{1}
}


%
\authorrunning{
}
%
\institute{Columbia University, Department of Computer Science, New York, NY, 10027 \and
Cornell University, School of Operations Research and Information Engineering, Ithaca, NY 14853 \and
Harvard University, Department of Systems Biology, Harvard Medical School, Boston, MA 02115 \and
Ben-Gurion University, Department of Computer Science, Israel, 8410501
}
\maketitle              
\begin{abstract}
Proteins are the major building blocks of life, and actuators of almost all chemical and biophysical events in living organisms. Their native structures in turn enable their biological functions which have a fundamental role in drug design. This motivates predicting the structure of a protein from its sequence of amino acids, a fundamental problem in computational biology. In this work, we demonstrate state-of-the-art protein structure prediction (PSP) results using embeddings and deep learning models for prediction of backbone atom distance matrices and torsion angles. We recover 3D coordinates of backbone atoms and reconstruct full atom protein by optimization. We create a new gold standard dataset of proteins which is comprehensive and easy to use. Our dataset consists of amino acid sequences, Q8 secondary structures, position specific scoring matrices, multiple sequence alignment co-evolutionary features, backbone atom distance matrices, torsion angles, and 3D coordinates. We evaluate the quality of our structure prediction by RMSD on the latest Critical Assessment of Techniques for Protein Structure Prediction (CASP) test data and demonstrate competitive results with the winning teams and AlphaFold in CASP13 and supersede the results of the winning teams in CASP12. We make our data, models, and code publicly available.

\keywords{Protein Structure Prediction \and Deep Learning \and Embeddings.}
\end{abstract}
\section{Introduction}
Proteins are necessary for a variety of functions within cells including transport, antibody, enzyme and catalysis. They are polymer chains of amino acid residues, whose sequences (aka primary structure) dictate stable spatial conformations, with particular torsion angles between successive monomers. To perform their functions, the amino acid residues must fold into proper configurations. The sequence space of proteins is vast, 20 possible residues per position, and evolution has been sampling it over billions of years. Thus, current proteins are highly diverse in sequences, structures and functions. The high throughput acquisition of DNA sequences, and therefore ubiquity of known protein sequences stands in contrast to the limited availability of 3D structures, which are more functionally relevant. 

From a physics standpoint, the process of protein folding is a search for the minimum energy conformation which happens in nature paradoxically fast \cite{levinthal1969how}. Unfortunately, explicitly computing the energy of a protein conformation, along with its surrounding water molecules, is extremely complex. Forward simulation approaches have been prohibitively slow on even modest-length proteins, with successes reported only for smaller peptides across timescales that are orders of magnitude too short, and even these required specialized hardware \cite{shaw_millisecond-scale_2009} or massive crowd sourcing \cite{larson2009folding}. Inferring local {\em secondary structure} \cite{drori2018q8} consists of linear annotation of structural elements along the sequence \cite{kabsch1983dictionary}. Inferring {\em tertiary structure} consists of resolving the 3D atom coordinates of proteins. When highly similar sequences are available with known structures, this homology can be used for modeling. Recently, PSP had been tackled by first predicting contact points between amino acids, and then leveraging the contact map to infer structure. A primary indicator of contact between a pair of amino acids is their tendency to have correlated and compensatory mutations during evolution. The availability of large scale data on DNA, and therefore protein sequences allows detection of such co-evolutionary constraints from sets of sequences that are homologous to a protein of interest. Registering such contacting pairs in a matrix facilitates a framework for their probabilistic prediction. This contact map matrix can be generalized to register distances between amino acids \cite{xu2019distance}. 

Machine learning approaches garnered recent success in PSP \cite{anand2018generative,alquraishi2019endtoend} and its sub-problems \cite{wang2018computational}. These leverage available repositories of tertiary structure \cite{pdb} and its curated compilations \cite{orengo1997cath,alquraishi2019proteinnet} as training data for models that predict structure from sequence. Specifically, the recent biannual critical assessment of PSP methods \cite{casp13} featured multiple such methods. Most prominently, a closed-source, ResNet-based architecture \cite{alphafold2019} has achieved impressive results in the CASP evaluation settings, based on representing protein structures both by their distance matrices as well as their torsion angles. 

\begin{figure}[t]
    \centering
    \includegraphics[width=1\linewidth]{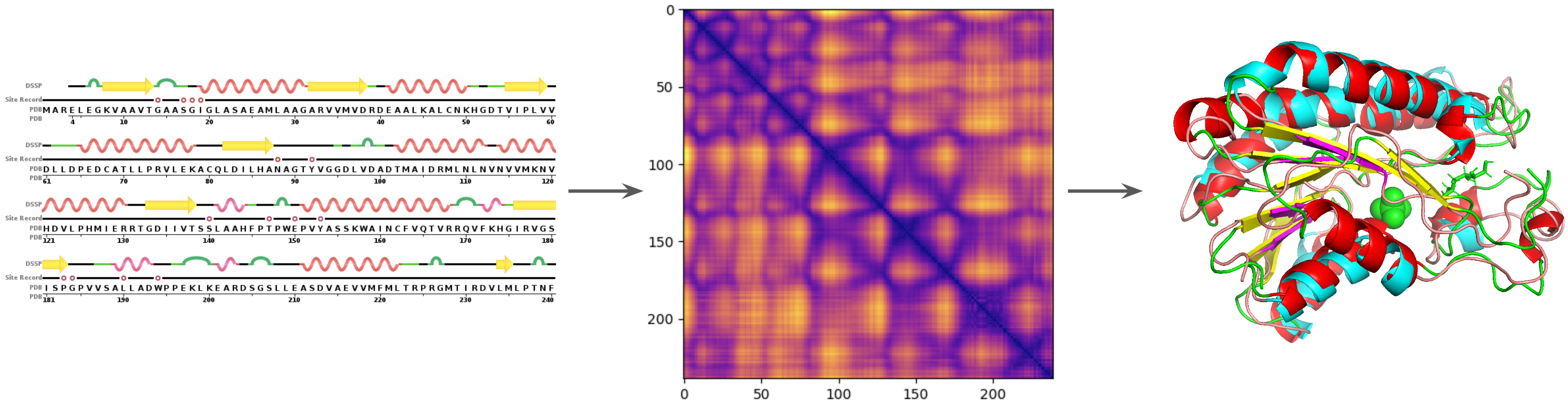}
    \caption{High level flow. The method operates in three stages by i) predicting backbone atom distance matrices and torsion angles ii) recovering backbone atom 3D coordinates, and iii) reconstructing the full atom protein by optimization}
    \label{fig:highlevelflow}
\end{figure}

In this work, we design a novel representation of biologically relevant input data and construct a processing flow for PSP as shown in Figure \ref{fig:highlevelflow}. Our method leverages advances in deep sequence models and proposes a method to learn transformations of amino acids and their auxiliary information. The method operates in three stages by i) predicting backbone atom distance matrices and torsion angles ii) recovering backbone atom 3D coordinates, and iii) reconstructing the full atom protein by optimization.

We demonstrate state-of-art protein structure prediction results using deep learning models to predict backbone atom distance matrices and torsion angles, recover backbone atom 3D coordinates and reconstruct the full atom protein by optimization. Our three key contributions are:
\begin{itemize}
    \item[$\bullet$] A gold standard dataset of around 75k proteins described in Table \ref{tab:dataset} which we call the CUProtein dataset which is easy to use in developing deep learning models for PSP.
    \item[$\bullet$] Competitive results with the winning teams on CASP13 and a comparison with AlphaFold (A7D) \cite{alphafold2019} and results mostly superseding the winning teams on CASP12.
    \item[$\bullet$] Publicly available source code for both protein structure prediction using deep learning models and protein reconstruction.
\end{itemize}
This work explores encoded representation for sequences of amino acids alongside their auxiliary information. We offer full access to data, models, and code, which remove significant barriers to entry for investigators and make publicly available methods for this important application domain.

\begin{figure}
\centering
\subfigure{
\includegraphics[width=0.26\textwidth]{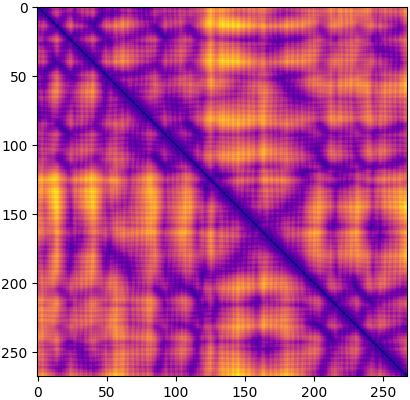}}
\subfigure{
\includegraphics[width=0.26\textwidth]{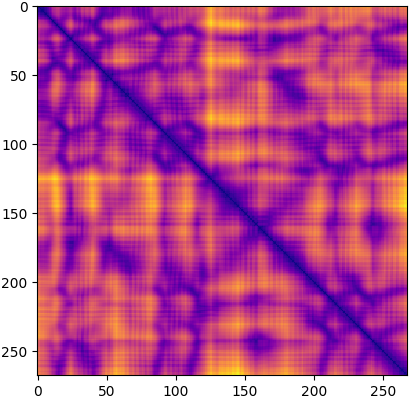}}
\subfigure{
\includegraphics[width=0.32\textwidth]{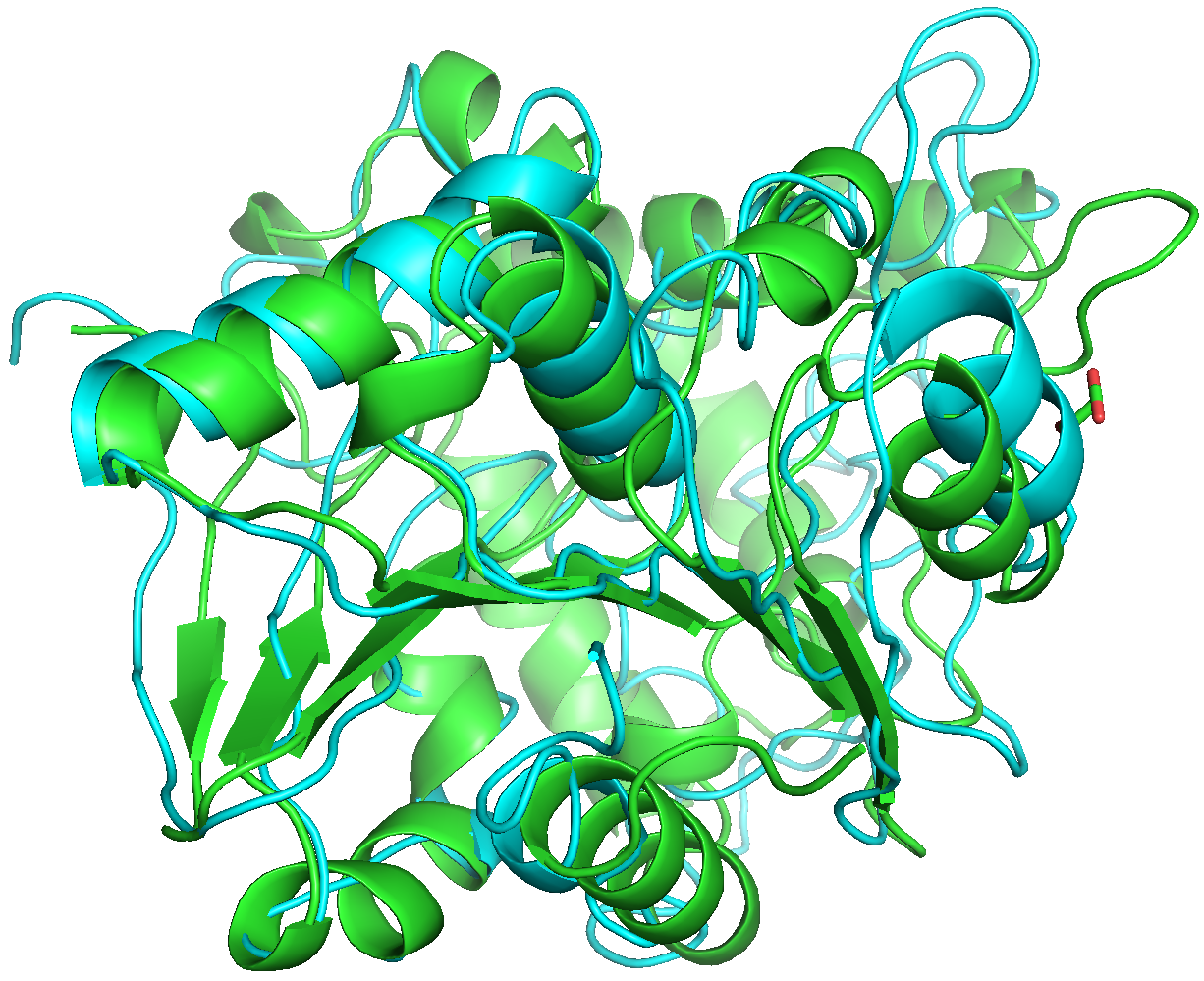}}
\subfigure{
\includegraphics[width=0.26\textwidth]{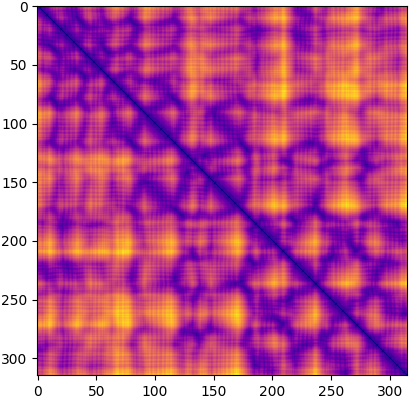}}
\subfigure{
\includegraphics[width=0.26\textwidth]{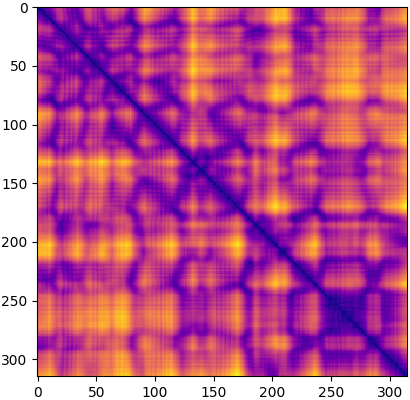}}
\subfigure{
\includegraphics[width=0.32\textwidth]{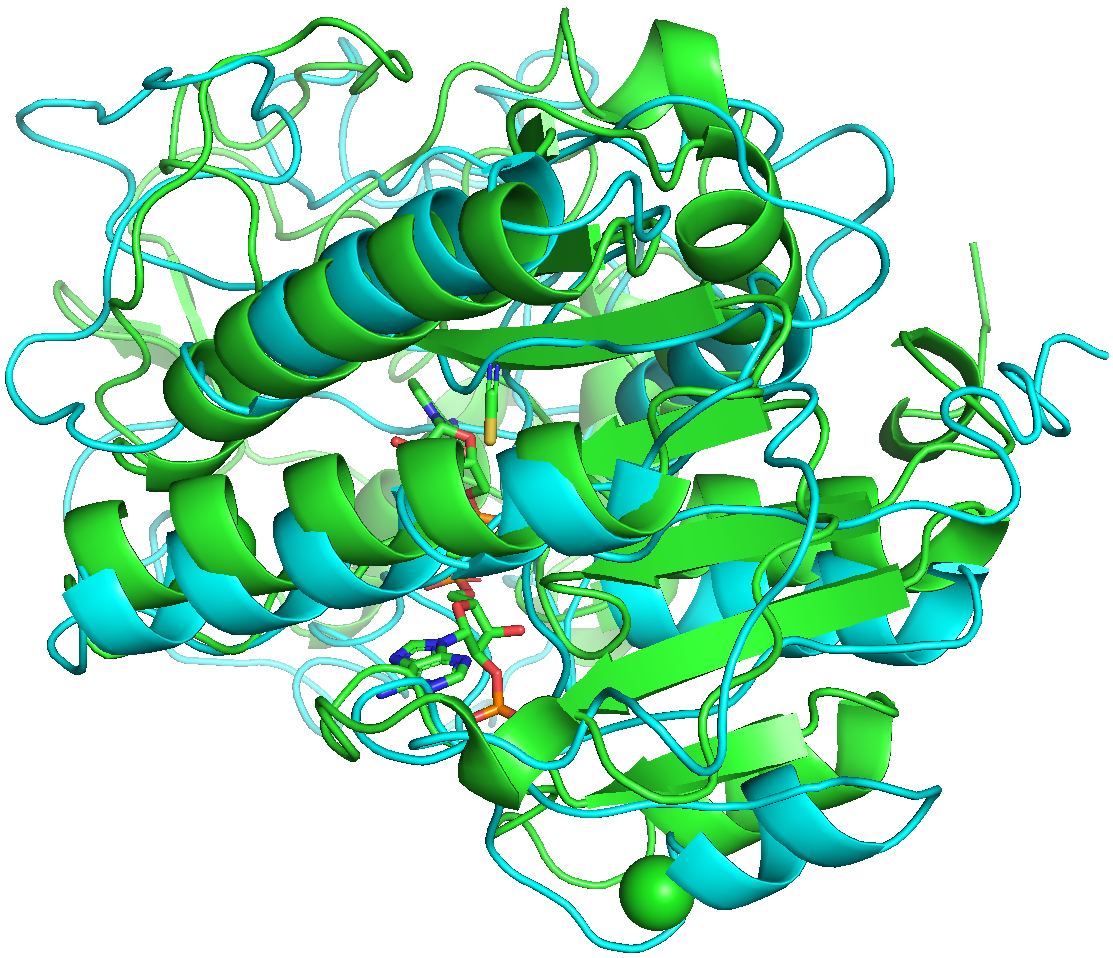}}
\setcounter{subfigure}{0}
\subfigure[]{
\includegraphics[width=0.26\textwidth]{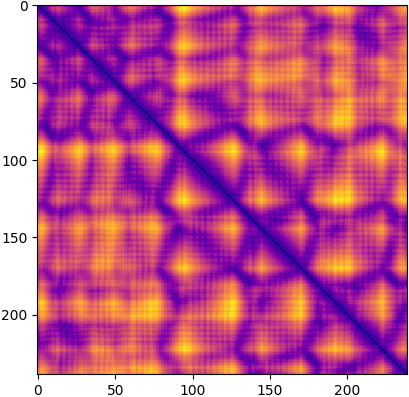}}
\subfigure[]{
\includegraphics[width=0.26\textwidth]{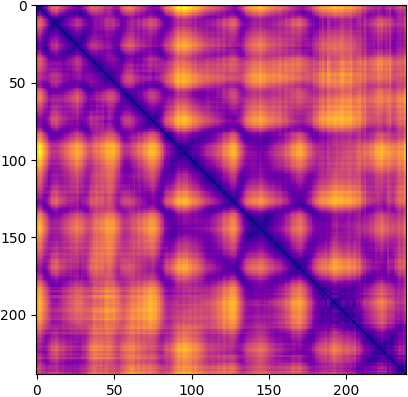}}
\subfigure[]{
\includegraphics[width=0.32\textwidth]{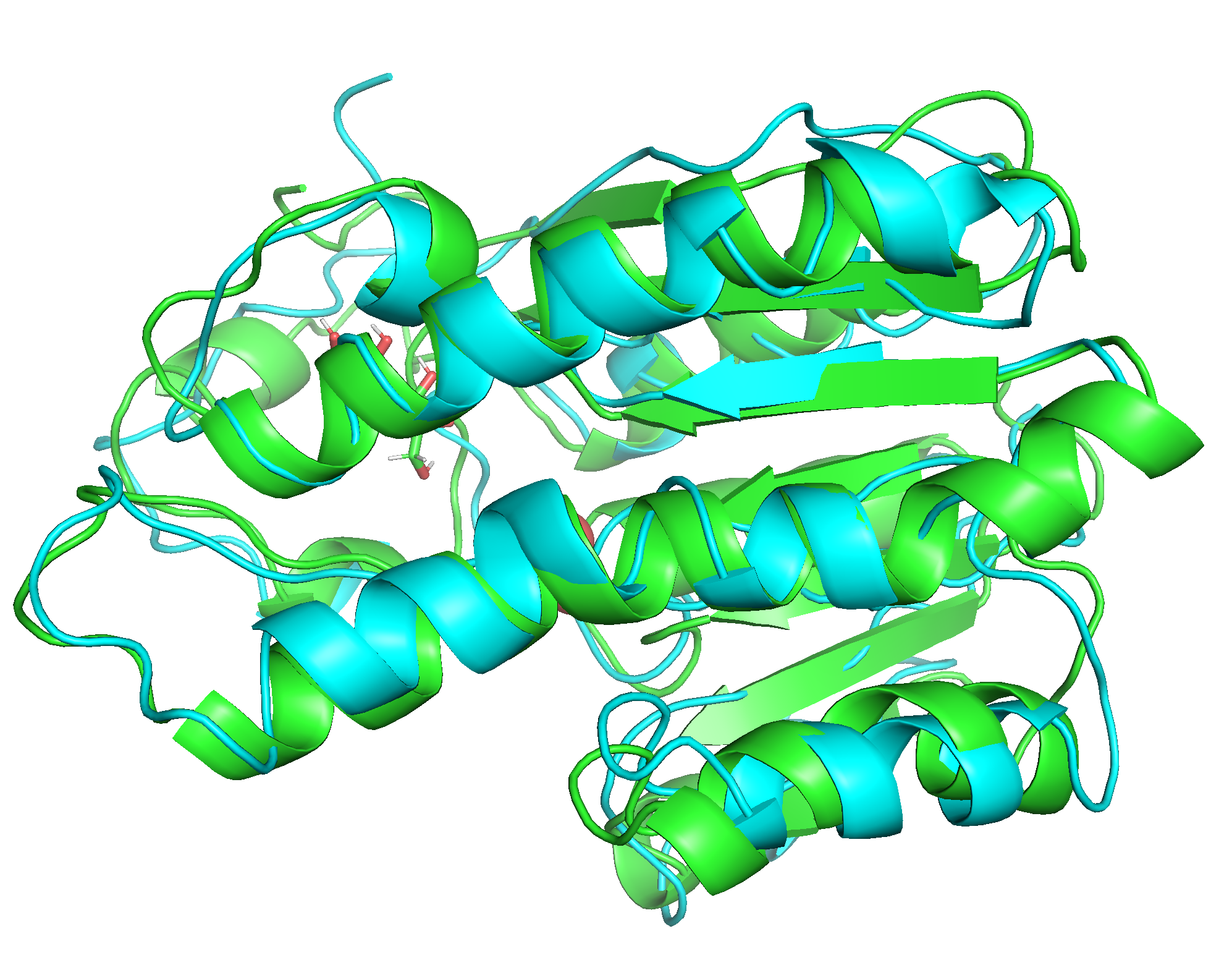}}
\caption{Visualization of CASP12 and CASP13 results: top row denotes protein 5Z82, center row protein 6D2V, and bottom row protein 5JO9. Columns denote (a) ground truth and (b) predicted  $C_\alpha$ distance matrices; (c) native (green) and predicted (cyan) structures using our model.}
\label{fig:sample}
\end{figure}

\section{Methods}
We address two problems: (i) predicting backbone distance matrices and torsion angles of backbone atoms from amino acid sequences, Q8 secondary structure, PSSMs, and co-evolutionary multiple sequence alignment; and (ii) reconstruction of all atom coordinates from the predicted distance matrices and torsion angles. Figure \ref{fig:sample} shows examples of ground truth backbone distance matrices and their predictions for CASP13 and CASP12 proteins and the corresponding full atom 3D native structures and reconstructed proteins.

\begin{table*}
\centering
\small
\begin{tabular}{lllllll} 
\textbf{Feature} & \textbf{Notation} & \textbf{Source} & \textbf{Dims} & \textbf{IO} & \textbf{Type} \\
\hline
AA Sequence & $x_{1}$ & PDB & $n \times 1$ & Input & $21$ chars \\ 
PSSM & $x_{2}$ & AA/HHBlits & $n \times 21$ & Input & Real [0,1]\\
MSA covariance & $x_{3}$ & AA/jackHMMER & $n \times n$ & Input & Real [0,1]\\ 
Secondary Structure (SS) & $x_{4}$ & DSSP & $n \times 8$ & Input &  8 chars\\
$C_\alpha$, $C_\beta$ Distance Matrices & $g_{1}(\cdot)$ & PDB & $n \times n$ & Output & \textup{\AA}ngstrom \\
Torsion Angles ($\phi$, $\psi$) & $g_{2}(\cdot)$ & DSSP & $n \times 2$ & Output & Radians\\
\hline
\end{tabular}
\caption{Our CUProtein dataset consists of amino acid sequences, secondary structure, PSSM's, MSA covariance matrices, backbone distance matrices, and torsion angles.}
\label{tab:dataset}
\end{table*}

\subsection{Predicting Backbone Distance Matrices and Torsion Angles}
Our inputs are AA sequences, Q8 sequences, PSSMs, and MSA covariance matrices. Our outputs are backbone distance matrices and torsion angles.

\paragraph{Embeddings.} We begin with a one-hot representation of each amino acid and secondary structure sequence, and real valued PSSM's and MSA covariance matrices. These are passed through \emph{embedding layers} and then onto an encoder-decoder architecture. To leverage sequence homology, we compute the covariance matrix of the MSA features by embedding the homology information along a $k$ dimensional vector to form a 3-tensor and contract the tensor $A_{ijk}$ along the $k$ dimensional embedding which is then passed as input to the encoder:
\begin{equation}
    \Sigma = A_{ji}^{k} A_{ijk},
\end{equation}

\begin{figure}[ht]
    \centering
    \subfigure[]{
    \includegraphics[width=0.46\linewidth]{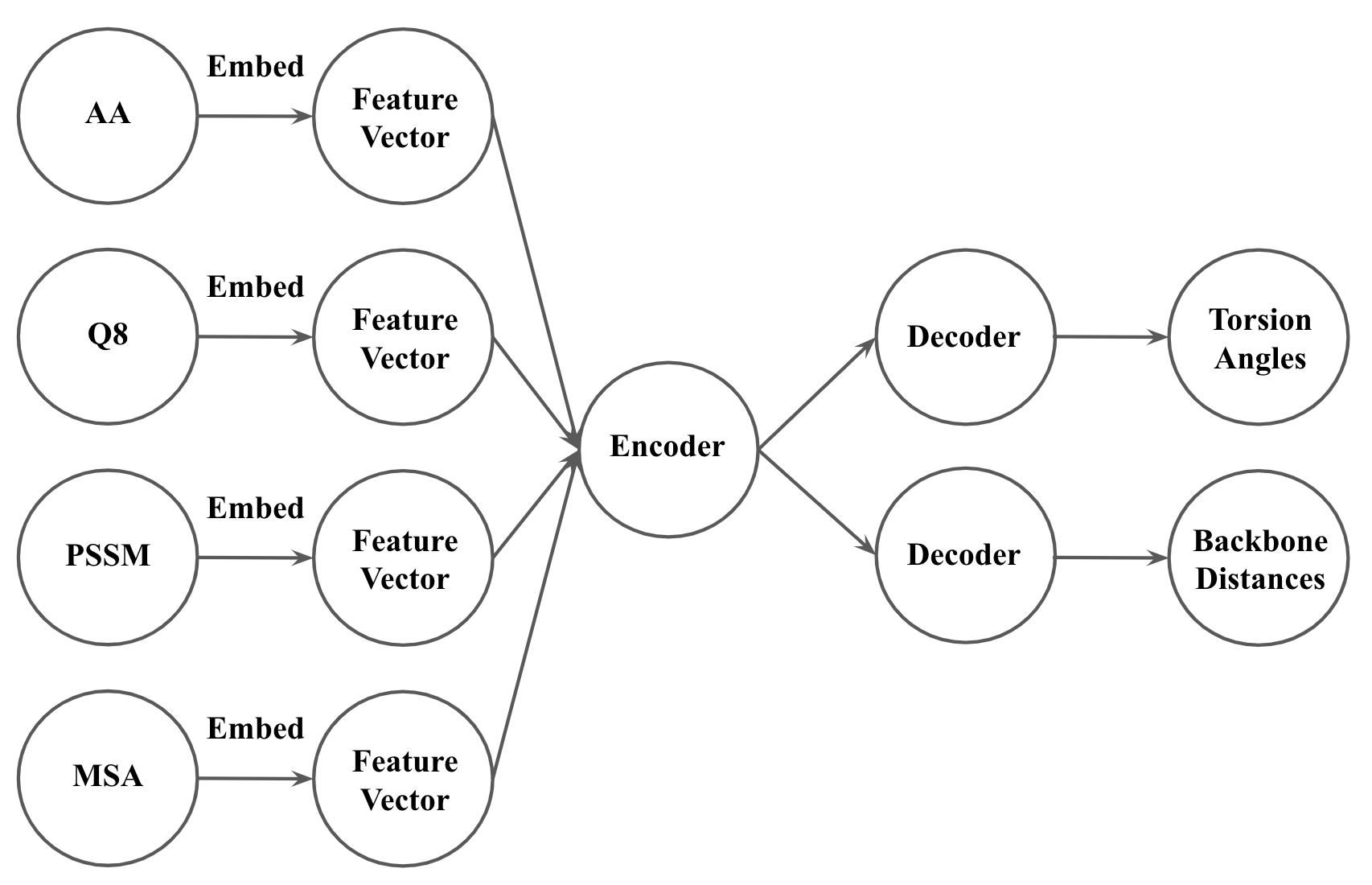}}
    \subfigure[]{
    \includegraphics[width=0.51\linewidth]{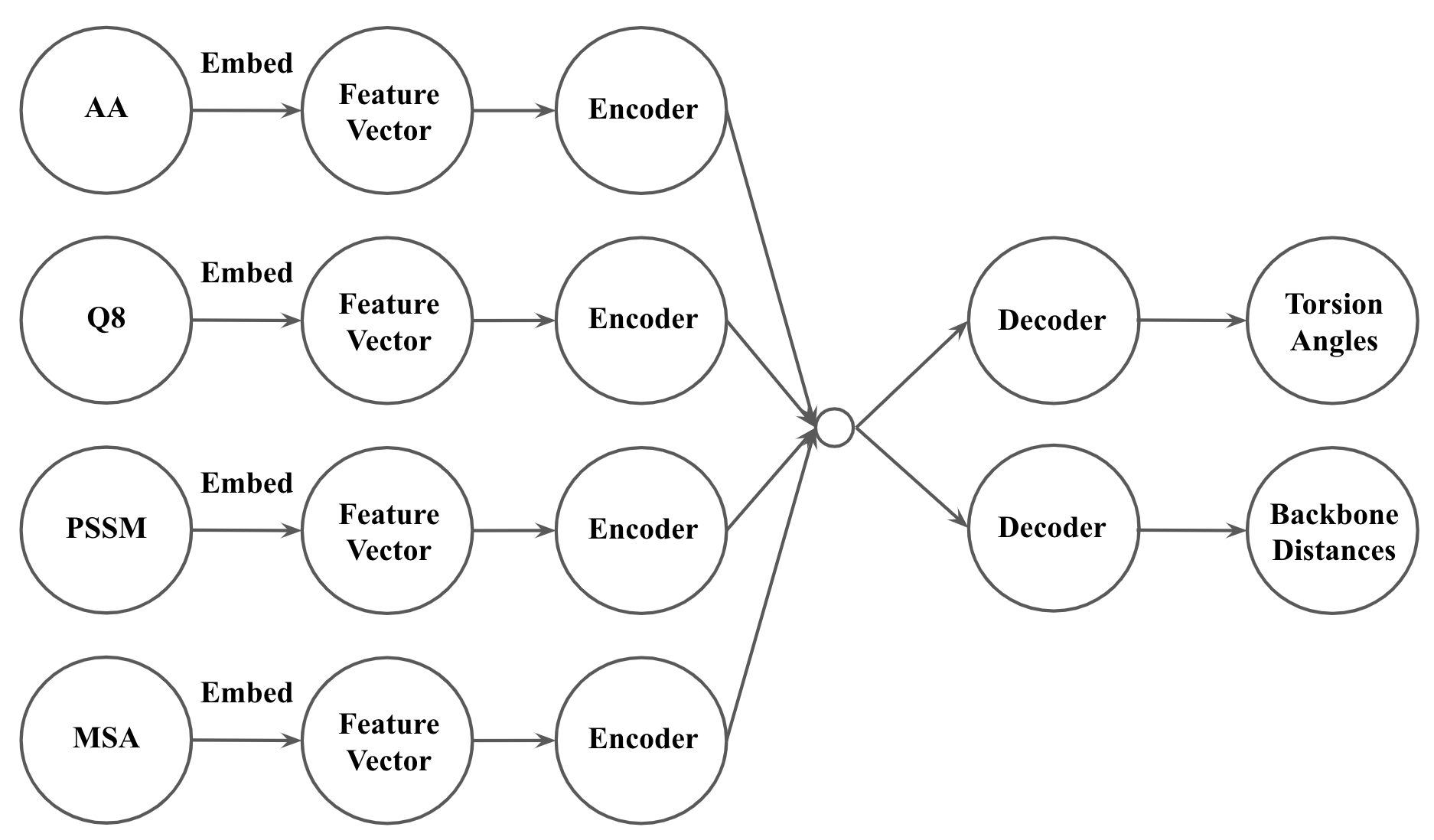}}
    \caption{Model architecture: inputs are embedded followed by (a) aggregation, encoding using sequence models, and decoding; (b) encoding using sequence models, aggregation, and decoding.}
    \label{fig:architecture}
\end{figure}

\paragraph{Encoder-Decoder architectures}
We use encoder-decoder models with a bottleneck to train prediction models. The encoder $f$ receives as input the aggregation $A$ (by concatenation) of the embeddings $e_{i}$ of each input $x_{i}$, and two separate decoders $g_{1}$ and $g_{2}$ that output distance matrices and torsion angles for $i\in\{1,\ldots,4\}$ as shown in Figure \ref{fig:architecture} (a). In addition, we also use a model which consists of separate encoders $f_{i}$ for each embedded input $e_{i}(x_{i})$, that are aggregated by concatenation after encoding, and separate decoders $g_{1}$ and $g_{2}$ for torsion angles and backbone distance matrices as shown in Figure \ref{fig:architecture} (b).
\begin{equation}
g_{j}\left(f\left(\underset{x_{i}\in{\mathcal{X}}}{\text{Agg}}(e_{i}(x_{i}))\right)\right),\
\quad
g_{j}\left(\underset{x_{i}\in{\mathcal{X}}}{\text{Agg}}\left(f_{i}(e_{i}(x_{1}))\right)\right). 
\end{equation}
Using a separate encoder model involves a larger number of trainable parameters. Our models differ in the use of embeddings for the input, their models, and loss functions.

\paragraph{Deep learning models} Building on techniques commonly used in natural language processing (NLP) our models use embeddings and a sequence of bi-directional GRU's and LSTM's with skip connections, and include batch normalization, dropout, and dense layers. We experimented with various loss functions including MAE, MSE, Frobenius norm, and distance logarithm to handle the dynamic range of distances. To learn the loss function we have also implemented distance matrix prediction using conditional GAN's and VAE's for protein sub-sequences.

\subsection{Reconstructing Proteins}
Once backbone distance matrices and torsion angle are predicted we address two reconstruction sub-problems: (i) reconstructing the protein backbone coordinates from their distance matrices, and (ii) reconstructing the full atom protein coordinates from the $C_{\alpha}$ or $C_{\beta}$ coordinates and torsion angles.

\paragraph{Backbone coordinate reconstruction.} We employ three different techniques for reconstructing the 3D coordinates $X$ between backbone atoms of a protein from the predicted matrix of their pairwise distances \cite{dokmanic2015euclidean}. Given a predicted distance matrix $\hat{D}$ our goal is to recover 3D coordinates $\hat{X}$ of $n$ points. We notice that $D(X)$ depends only on the Gram matrix $X^T X$:
\begin{equation}
	D(X) = \mathbf{1} \text{diag}(X^T X)^T - 2 X^T X + \text{diag}(X^T X) \mathbf{1}^T,
\end{equation}

\emph{Multi-dimensional scaling (MDS):}
\begin{equation} \label{eq:mds}
	\underset{\hat{X}}{\text{minimize}} \norm{D(\hat{X}) - \hat{D}}{F}^2.
\end{equation}
\emph{Semi-definite programming (SDP) and relaxation (SDR):}
 \begin{equation} 
    	\underset{G}{\text{minimize}} \norm{K(G) - \hat{D}}{F}^2 \label{eq:sd} \qquad
	\text{s.t.} \ \ \ \
	 G \in C,
\end{equation}
where $K(G) = \mathbf{1} \text{diag}(G)^T - 2 G + \text{diag}(G) \mathbf{1}^T$ operates on the Gram matrix $G$.

\emph{Alternating direction method of multipliers (ADMM) \cite{anand2018generative}:}
\begin{align} 
\begin{aligned}
    \label{eq:admm}
	\underset{G, Z, \eta}{\text{minimize}} \ \ \lambda \norm{\eta}{1} &+ \frac{1}{2} \left( \sum_{i,j = 1}^n  (G_{ii} + G_{jj} - 2 G_{ij} + \eta_{ij} - \hat{D}_{ij}^2) \right)^2 \nonumber \\
	&+ \mathbbm{1} \{Z \in S_+^n\} \ \ \text{s.t.} \ \  G - Z = 0.
\end{aligned}
\end{align}
We have found multi-dimensional scaling to be the fastest and most robust method of the three, without depending on algorithm hyper-parameters, which is most suitable for our purposes. To handle the chirality of these coordinates we calculate all alpha-torsion angles (the torsion angles defined by four consecutive $C_{\alpha}$ atoms) and compare their distribution to the characteristic, highly non-symmetric distribution of protein alpha-torsions. If the predicted alpha-torsions are unlikely to result from the distribution of the protein dataset, we invert the coordinates chirality.

\paragraph{Full atom coordinate reconstruction.} Given backbone coordinates we assign plausible coordinates to the rest of the protein's atoms. We begin with an initial guess or prediction for $\phi$ and $\psi$ torsion angles. We maintain a look-up table of mean $\phi$, $\psi$ values for each combination of two consecutive $\alpha$ torsions and three $\alpha$-angles (the angles defined by three consecutive atoms). Using these values and the $C_{\alpha}$ positions we generate an initial model. This model is then relaxed by a series of energy minimization simulations under an energy function that includes: standard bonded terms (bond, angle, plane and out-of-plane), knowledge based Ramachandran and pairwise terms, torsion constraints on the $\phi$ and $\psi$ angles, and tether constraints on the $C_{\alpha}$ position. The latter term reduces the perturbations of the initial high forces. Finally, we add side-chains using a rotamer library, and remove clashes by a series of energy minimization simulations. We develop a very similar method for reconstructing the full atom protein from $C_{\beta}$ atom distance matrices.

\section{Results}

\begin{table*}
\centering
\small
\begin{tabular}{cccccc}
\textbf{PDB} & \textbf{CASP} & \textbf{Tar-id} & \textbf{Best RMSD} & \textbf{A7D} & \textbf{Ours}\\
\hline
5Z82 & 13 & T0951 & 1.01 (Seok) & NA & 1.79\\
6D2V & 13 & T0965 & 1.72 (A7D) & 1.72 & \textbf{1.60}\\
6QFJ & 13 & T0967 & 1.13 (BAKER) & NA & 1.18\\
6CCI & 13 & T0969 & 1.96 (Zhang) & 2.27 & 2.53\\
6HRH & 13 & T1003 & 0.88 (MULTICOM) & 2.12 & 2.95\\
6QEK & 13 & T1006 & 0.58 (YASARA) & 0.78 & 1.02\\
6N91 & 13 & T1018 & 1.24 (Wallner) & 1.77 & 3.89\\
6M9T & 13 & T1011-D1 & 1.58 (A7D) & 1.58 & 1.64\\
\hline
5J5V & 12 & T0861 & 0.49 (MULTICOM) & NA & 1.00\\ 
2N64 & 12 & T0865 & 1.87 (HHPred) & NA & \textbf{1.58}\\ 
5JMU & 12 & T0879 & 1.35 (MULTICOM) & NA & \textbf{1.31}\\ 
5JO9 & 12 & T0889 & 1.31 (Seok) & NA & 1.79\\ 
4YMP & 12 & T0891 & 1.10 (GOAL) & NA & 1.36\\ 
5XI8 & 12 & T0942-D2 & 1.73 (EdaRose) & NA & \textbf{1.60}\\ 

\hline
\end{tabular}
\caption{Comparison between the winning CASP12/13 models for each protein, best AlphaFold (A7D) model for CASP13, and our model for CASP12/13.}
\label{tab:results}
\end{table*}

We have compared our predictions on CASP12 and CASP13 \cite{abriata2019further} test targets. Deep learning methods, including were widely used only starting from CASP13. AlphaFold was introduced starting from CASP13. The use of deep learning methods in CASP13, due to availability of DL programming frameworks, significantly improved performance over CASP12. Table \ref{tab:results} provides a representative comparison between the winning CASP12 and CASP13 competition models, AlphaFold models for CASP13 for which A7D submitted predictions to CASP, and our models. These proteins were selected such that they do not contain multiple chains and are sufficiently complex. The average length of the CASP13 target proteins selected is 325 residues, and the average for CASP12 is 247 residues. Results of RMSD around 2 Angstrom on test targets are considered accurate in CASP.

Our results supersede the winning teams of CASP12 compared with each best team for each protein which highlights the improvement using deep learning methods. Our approach is on par with the winning teams in CASP13, compared with the winning team for each protein, which highlights that our methos is state-of-the-art. We measure the sequence independent RMSD which is consistent with the CASP evaluation reports and matches the deposited structures and our predictions. CASP competitors such as AlphaFold provide predictions for selected proteins. Overall, our performance on CASP is highly competitive. Training our models on a cloud instance takes two days using GPUs. Prediction of backbone distance matrices and torsion angles takes a few seconds per protein, and reconstruction of full atom proteins from distance matrices and torsion angles takes a few minutes per protein, depending on protein length. Limitations of this work are: (i) we only handle single domain proteins and not complexes with multiple chains, (ii) PSSM and MSA data for several of the CASP targets are limited to a subsequence of the full length protein, and (iii) we do not use available methods for detecting and reconstructing beta-sheets.

\section{Conclusions}
We present novel deep learning representations and models for protein structure prediction. We provide a new and open dataset which can easily be used, and make our models and code publicly available \cite{drori19cutsprepo}. We predict accurate structures around 2 Angstrom within the native structures. Our results are competitive with the latest CASP13 and AlphaFold results and supersede CASP12 results.

\bibliography{bibliography}

\begin{thebibliography}{10}

\bibitem{abriata2019further}
Luciano~A Abriata, Giorgio~E Tam{\`o}, and Matteo Dal~Peraro.
\newblock A further leap of improvement in tertiary structure prediction in
  {CASP13} prompts new routes for future assessments.
\newblock {\em Proteins: Structure, Function, and Bioinformatics}, 2019.

\bibitem{alquraishi2019endtoend}
Mohammed AlQuraishi.
\newblock End-to-end differentiable learning of protein structure.
\newblock {\em Cell systems}, 8(4):292--301, 2019.

\bibitem{alquraishi2019proteinnet}
Mohammed AlQuraishi.
\newblock Protein{N}et: a standardized data set for machine learning of protein
  structure.
\newblock {\em BMC bioinformatics}, 20(1):311, 2019.

\bibitem{anand2018generative}
Namrata Anand and Possu Huang.
\newblock Generative modeling for protein structures.
\newblock In {\em Advances in Neural Information Processing Systems}, pages
  7494--7505, 2018.

\bibitem{casp13}
Protein Structure~Prediction Center.
\newblock 13th community wide experiment on the critical assessment of
  techniques for protein structure prediction, 2018.

\bibitem{alphafold2019}
Google DeepMind.
\newblock Alpha{F}old: De novo structure prediction with deep-learning based
  scoring, 2018.

\bibitem{dokmanic2015euclidean}
Ivan Dokmanic, Reza Parhizkar, Juri Ranieri, and Martin Vetterli.
\newblock Euclidean distance matrices: essential theory, algorithms, and
  applications.
\newblock {\em IEEE Signal Processing Magazine}, 32(6):12--30, 2015.

\bibitem{drori2018q8}
Iddo Drori, Isht Dwivedi, Pranav Shrestha, Jeffrey Wan, Yueqi Wang, Yunchu He,
  Anthony Mazza, Hugh Krogh-Freeman, Dimitri Leggas, Kendal Sandridge, et~al.
\newblock High quality prediction of protein q8 secondary structure by diverse
  neural network architectures.
\newblock {\em NeurIPS Workshop on Machine Learning for Molecules and
  Materials}, 2018.

\bibitem{drori19cutsprepo}
Iddo Drori, Darshan Thaker, Arjun Srivatsa, Daniel Jeong, Yueqi Wang, Linyong
  Nan, Fan Wu, Dmitri Leggas, Jinhao Lei, Weiyi Lu, Weilong Fu, Yuan Gao,
  Sashank Karri, Anand Kannan, Antonio Moretti, Mohammed AlQuraishi, Chen
  Keasar, and Itsik Pe'er.
\newblock Git{H}ub repository for {A}ccurate protein structure prediction by
  embeddings and deep learning representations.
\newblock \url{https://github.com/idrori/cu-tsp}, 2019.

\bibitem{pdb}
Z.~Feng G. Gilliland T.N. Bhat H. Weissig I.N. Shindyalov P.E.~Bourne
  H.M.~Berman, J.~Westbrook.
\newblock The protein data bank.
\newblock {\em Nucleic Acids Research}, 28, 2000.

\bibitem{kabsch1983dictionary}
Wolfgang Kabsch and Christian Sander.
\newblock Dictionary of protein secondary structure: pattern recognition of
  hydrogen-bonded and geometrical features.
\newblock {\em Biopolymers: Original Research on Biomolecules},
  22(12):2577--2637, 1983.

\bibitem{larson2009folding}
Stefan~M Larson, Christopher~D Snow, Michael Shirts, and Vijay~S Pande.
\newblock Folding@ home and genome@ home: Using distributed computing to tackle
  previously intractable problems in computational biology.
\newblock {\em arXiv preprint arXiv:0901.0866}, 2009.

\bibitem{levinthal1969how}
Cyrus Levinthal.
\newblock How to fold graciously.
\newblock In {\em Proceedings of a meeting held at Allerton House. P.
  Debrunner, JCM Tsibris, and E. Munck, editors. University of Illinois Press,
  Urbana, IL}, 1969.

\bibitem{orengo1997cath}
Christine~A Orengo, Alex~D Michie, Susan Jones, David~T Jones, Mark~B
  Swindells, and Janet~M Thornton.
\newblock Cath--a hierarchic classification of protein domain structures.
\newblock {\em Structure}, 5(8):1093--1109, 1997.

\bibitem{shaw_millisecond-scale_2009}
David~E. Shaw, Ron~O. Dror, John~K. Salmon, J.~P. Grossman, Kenneth~M.
  Mackenzie, Joseph~A. Bank, Cliff Young, Martin~M. Deneroff, Brannon Batson,
  Kevin~J. Bowers, Edmond Chow, Michael~P. Eastwood, Douglas~J. Ierardi,
  John~L. Klepeis, Jeffrey~S. Kuskin, Richard~H. Larson, Kresten
  Lindorff-Larsen, Paul Maragakis, Mark~A. Moraes, Stefano Piana, Yibing Shan,
  and Brian Towles.
\newblock Millisecond-scale {Molecular} {Dynamics} {Simulations} on {Anton}.
\newblock In {\em Proceedings of the {Conference} on {High} {Performance}
  {Computing} {Networking}, {Storage} and {Analysis}}, {SC} '09, pages
  39:1--39:11, 2009.

\bibitem{wang2018computational}
Jingxue Wang, Huali Cao, John~ZH Zhang, and Yifei Qi.
\newblock Computational protein design with deep learning neural networks.
\newblock {\em Scientific reports}, 8(1):6349, 2018.

\bibitem{xu2019distance}
Jinbo Xu.
\newblock Distance-based protein folding powered by deep learning.
\newblock {\em Proceedings of the National Academy of Sciences},
  116(34):16856--16865, 2019.

\end{thebibliography}
\bibliographystyle{plain}
\end{document}